# Ground state of bond-disordered quasi-one-dimensional spin system $(CH_3)_2CHNH_3Cu(Cl_xBr_{1-x})_3$ with $x$ = 0, 0.25 and 0.3

Takayuki GOTO[1], Takao SUZUKI[2], Isao WATANABE[3], Hirotaka MANAKA[4], Hubertus Luetkens[5] and Alex Amato[5]

[1]*Physics Division, Sophia University, 7-1 Kioi-cho, Chiyoda-ku, Tokyo 102-8554*
[2]*Faculty of Engineering, Shibaura Institute of Technology, Saitama 337-8570, Japan*
[3]*Advanced Meson Science Laboratory, RIKEN 2-1 Hirosawa, Wako, Saitama 351-0198*
[4] Graduate School of Science and Engineering, Kagoshima University, Kagoshima 890-0065
[5] Laboratory for Muon-Spin Spectroscopy, Paul Scherrer Institute, CH 5232 Villigen PSI, Switzerland

*E-mail: gotoo-t@sophia.ac.jp*

The ground state of the quasi-one-dimensional system with bond-disorder $(CH_3)_2CHNH_3Cu(Cl_xBr_{1-x})_3$ with $x$ = 0, 0.25 and 0.3 has been investigated by μSR. The Fourier spectrum of electron spin fluctuation in $x$ =0.25 and 0.3 obtained by LF-μSR technique shows that there exists soft-mode toward a possible phase transition to an exotic phase such as Bose-glass. In gapped system with $x$ = 0, the Fourier spectrum is totally different from the other two, indicating the existence of the quantum critical point at a finite $x$ between 0 and 0.25.

## 1. Introduction

The $S$=1/2 quasi-one-dimensional spin system $(CH_3)_2CHNH_3CuBr_3$, which is best described as an asymmetric quantum spin ladder, shows a singlet ground state with an energy gap of Δ=98 K [1,2,3]. With substituting Br halogen ion with Cl, the bond disorder is introduced and a new-type ground state is expected [3,4]. The measurements of macroscopic quantities in $(CH_3)_2CHNH_3Cu(Cl_xBr_{1-x})_3$ with $x$ = 0 – 0.44 have been intensively studied by H. M., who has concluded that the ground state at finite $x$ is anomalous, because the temperature dependence of the specific heat and the magnetic susceptibility suggests that the system remains to be gapped against Cl substitution until $x$ =0.44, while on the other hand, the magnetization curve shows a finite gradient [2,4].

The elementary spin excitation in this system is the triplet excitation, which can be considered as mobile boson particles in vacuum or the single sea. Fisher *et al.* theoretically studied the disordered lattice boson system with random onsite chemical potential to predict the existence of an exotic ground-state Bose glass [6,7,8]. The phase exists only at absolute zero and is characterized mostly by the property that bosons are localized, although they are massless.

The motivation of our study is to investigate microscopically the ground state of the mixture of these two nonmagnetic compounds. The mixture can be considered as the quantum spin system with disordered bonds. The disorder in this system differs qualitatively from the Mg-doped spin-Peierls systems such as $CuGeO_3$ in that there are no unpaired spins in the present system but only a random modulation of the bonds [9].

So far, we have been studied the disordered system by microscopic tools of LF-μSR and NMR [10,11,12] to reveal the existence of soft-mode toward absolute zero in the Fourier spectrum of spin fluctuation in the sample of $x$=0.35 [13] and 0.40 [14]. For the latter, in particular, we have shown that no magnetic order takes place in temperatures down to 15 mK indicating the realization of the exotic ground state with zero critical temperature. In this article, as a next step, we extend the investigation to a smaller range of $x \leq 0.3$ so as to investigate whether or not the quantum critical point exists in between the exotic phase and the gapped end member $x = 0$.

## 2. Experimental

Single crystals of $(CH_3)_2CHNH_3Cu(Cl_xBr_{1-x})_3$ with $x$ = 0, 0.25, 0.3 have been grown by a solvent evaporation method from an isopropylalcohol solution of $(CH_3)_2CHNH_2$-HX and $CuX_2$ (X = Cl, Br). As-grown crystals are black in color, flat and rectangular in shape [2,5], and typically 2×3×1 $mm^3$ in size. μSR measurements were performed on typically about 15 peaces of single crystal at the RIKEN-RAL Muon Facility in the UK and the Swiss Muon Source (SmS), PSI in Switzerland. The incident muon-spin direction was parallel to the $b$*-axis[1,2,10,13]. The decay curve of muon spin polarization was obtained from the ratio of the numbers of muon events counted by forward and backward counters; it is represented by the asymmetry

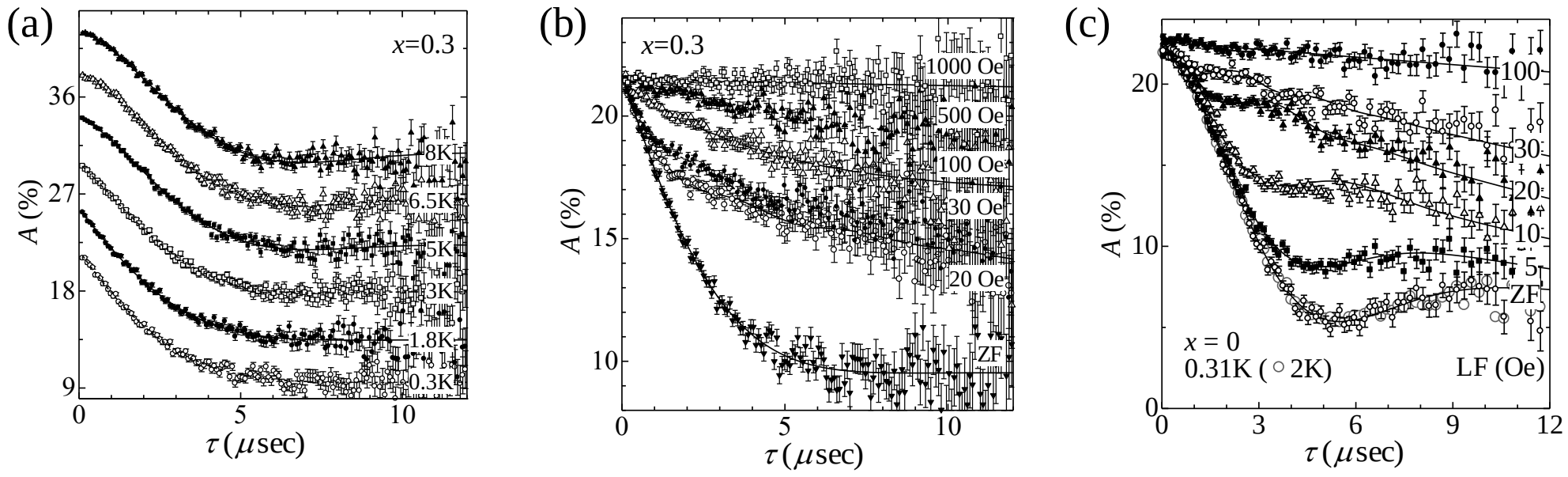

**Fig. 1.** Typical depolarization curves at various temperatures and the longitudinal fields (LF), (a) ZF spectra of $x$ = 0.3, (b) LF decoupling spectra for $x$=0.3 and (c) LF decoupling spectra for $x$ = 0, where solid curves show the fitting curves described in main text.

parameter $A(t)$.

## 3. Results

Typical relaxation curves of muon spin polarization under zero and finite LFs are shown in Fig. 1. No muon spin rotation due to a static hyperfine field was observed for all the samples investigated down to 0.3 K. This is consistent with the results of macroscopic quantities showing no indication of the magnetic order [13,14]. As the temperature was lowered, one can see that the depolarization becomes faster and that the shape of the short-time region of the curves changes from Gaussian-type to Lorentzian-type. Under the finite longitudinal field, depolarization curves of $x$ = 0.25 and 0.30 bear the two components of fast and slow rates, which can be seen apparently in Fig. 1 (b) for $x$ = 0.30. The observed depolarization curves were described by a function containing two components expressed as $A(t)=G_{\mathrm{KT}}(t,\Delta)\cdot\left(A_1e^{-\lambda_1 t}+A_2e^{-\lambda_2 t}\right)$ where $G_{\mathrm{KT}}$ is the Kubo-Toyabe function with parameters of a static field distribution width $\Delta$, depolarization rates of Lorentz-type $\lambda_1$ and $\lambda_2$, and component amplitudes $A_1$ and $A_2=1-A_1$. In determining these parameters, we first set all of them free and performed fitting to find that the quasi static nuclear part width $\Delta$ stays nearly constant to be 0.24(2) for $x$ = 0.25 and 0.23(2) for $x$ = 0.30. Next, we fixed $\Delta$ to be the averaged value 0.23($\mu s^{-1}$) and performed all the fitting again to obtain the other parameters [13,14].

As for the volume fraction of each component, we first set $A_1$ and $A_2$ as free and confirmed that the fraction does not vary against LF below 100 Oe. Above 100 Oe, where the slower depolarization rate $\lambda_1$ becomes negligibly small and hence contribute little to the depolarization curves of measured time range, we fixed the fraction to the low-field values and $\lambda_1$ to zero and assumed that the fractions does not change from the value taken at low fields. Observed depolarization curves above 50 Oe showed a good fit with the function based on this assumption. The volume fraction of $\lambda_2$ phase, that can be written as $A_2$ starts from 0.2 at 6.5 K to increase and reaches 0.4 at 0.3 K for both the samples of $x$ = 0.25 and 0.30. This behavior is commonly observed irrespective of $x$ in the region between 0.25 and 0.40 [13,14]. The temperature dependence of the two depolarization rates $\lambda_1$ and $\lambda_2$ under various LFs are shown in Fig. 2 (a) and (b).

The spin depolarization in $x$ = 0 bears only a single component with $\Delta$ = 0.32(2) in entire temperature and field region as shown in Fig. 1 (c). Its time evolution was well fitted by a single exponential as $A(t) = G_{KT}(t,\Delta)\cdot e^{-\lambda t}$. The LF dependence of the Lorentzian part of depolarization rate λ at 0.3 and 1.8 K thus obtained is shown and compared with $\lambda_1$ in the other samples in Fig. 2 (c). Generally, the decrease in λ with increasing LF maps the Fourier spectrum of the fluctuation of electron spins [13,15,16]. The disordered samples showed the Lorentz type spectrum, which suggests the electron spins in the classical paramagnetic state. For the $x$ = 0 sample, on the contrary, the Fourier spectrum was described as the sum of the Lorentz-type component and a white component with an amplitude of 11 %. The amplitude of the white component does not change with increasing temperature up to 1.8 K. By fitting to the so-called Redfield function[13,15,16], we obtained the parameters of the amplitude of the fluctuating local field $\delta H$ and the characteristic fluctuating frequency to find that they are nearly the same irrespective of $x$; at 0.3 K, $\delta H$ = 2.67 Oe, $\nu_0$ = 0.164 MHz for both $x$ = 0.25 and 0.30, and $\delta H$ = 3.56 Oe, $\nu_0$= 0.370 MHz for $x$ = 0.

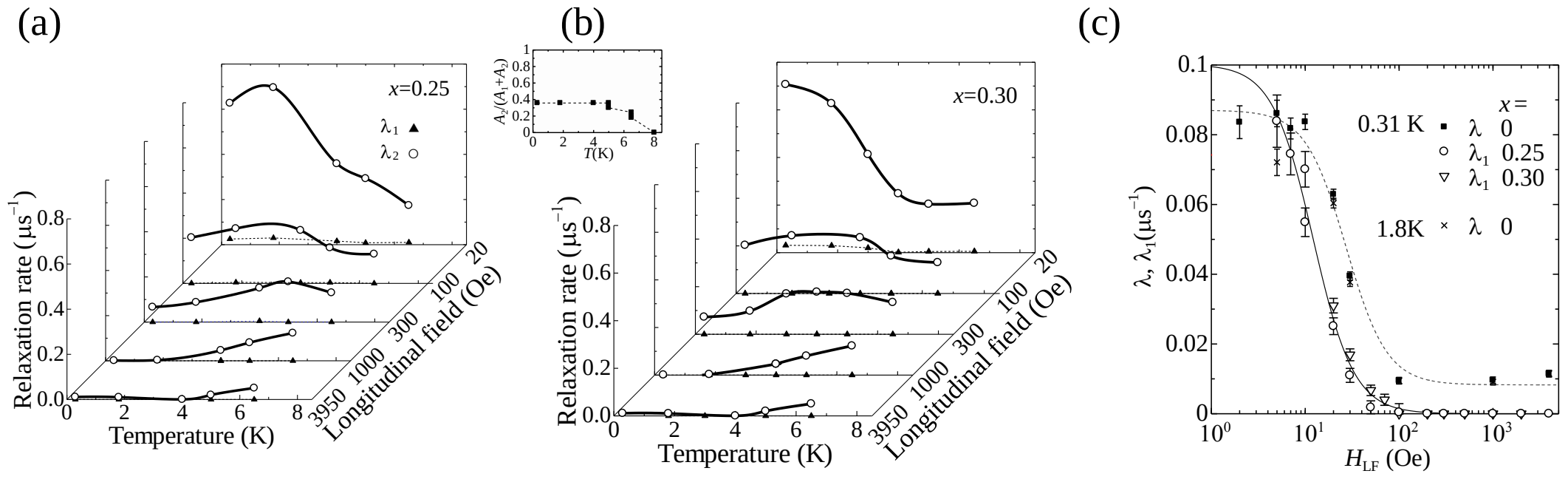

**Fig. 2.** (a), (b) Temperature and LF dependence of $\lambda_1$ and $\lambda_2$ for the samples $x$ = 0.25 and 0.30. Curves are guides to the eye. (c) LF-Decoupling spectra at 0.31 K for the samples $x$ = 0, 0.25 and 0.30. The solid and dashed curves show the Redfield function described in the text.

## 3. Discussion

First, we note that the existence of the two depolarization rate $\lambda_1$ and $\lambda_2$ indicates that the system with the bond disorder phase separates into the two components. This has been already reported in higher $x$ region, and interpreted as that active spins, described by $\lambda_2$, with the strong antiferromagnetic correlation are restricted in nano-size islands that are surrounded by the singlet sea described by $\lambda_1$ [13,14]. The LF dependence of the $\lambda_2$ in Fig. 2 shows that the Fourier spectrum of the spin fluctuation changes from white noise-like one at high temperature around 6 K to the steep peak weighed at around zero energy in low temperatures. This behavior is almost the same as that observed in $x$ = 0.35 and 0.40, and is interpreted to be the soft-mode toward the possible phase transition with zero $T_C$, such as Bose-glass [13,14].

Next, we move on to the $x$ = 0, where the Fourier spectrum of electron spin fluctuation consists of two components: the Lorentz-shaped spectrum and the white spectrum. The fact that both $\delta H$ and $\nu_0$ are nearly the same as those for the $\lambda_1$ components in $x$ = 0.25 and 0.30 confirms our idea that the component described by $\lambda_1$ inherits the character of the siglet phase, and that the other component described by $\lambda_2$ newly appears by the bond disorder. With decreasing $x$ toward zero, we expect that at the volume fraction of the latter vanishes and the white part of the spectrum appears in turn at the QCP. At present, we can only conclude that $x_{QCP}$ is much smaller than 0.25. In order to determine $x_{QCP}$ as well as the origin of the white spectrum observed only in $x$ = 0, an investigation on the still smaller $x$ region is necessary, which is now in preparation.

Finally, we refer to the effect of the muon spin disturbance to the singlet dimer. That is, λ in gapped systems often shows a quite anomalous dependence on LF at around the temperature comparable with Δ [17,18,19]. In our $x$ = 0 system, however, we did not observed any anomalous behavior of λ in the entire experimental temperature range. This may be due to the large spin gap of $\Delta \approx 98$ K in this system.

## 4. Conclusion

The Fourier spectrum of the electron spin fluctuation in quasi-one-dimensional quantum spin system with and without the bond-disorder was investigated by μSR. In disordered system, the spectrum showed the steep peak weighed at around zero energy in low temperatures, indicating the existence of soft-mode toward the possible phase transition with zero $T_C$, such as Bose-glass. In $x = 0$ system, the profile of spectra is totally different from disordered systems, indicating the existence of QCP at a finite $x$.

## Acknowledgment

This work was partially supported by JSPS KAKENHI Grant Number 21110518, 24540350 and 21540344.